\documentclass[conference]{IEEEtran}

\ifCLASSINFOpdf
\else
   \usepackage[dvips]{graphicx}
\fi

\usepackage{boldline,multirow}
\usepackage{cite}
\usepackage{amsmath,amssymb,amsfonts}
\usepackage{algorithmic}
\usepackage{algorithm}
\usepackage{graphicx}
\usepackage{textcomp}
\usepackage{xcolor}
\usepackage{booktabs}
\usepackage{amsmath}
\usepackage{array} 
\usepackage{tikz}
\usetikzlibrary{arrows.meta, positioning}
\usepackage{bm}
\usepackage{breakurl}
\usepackage{balance}
\usepackage{mathrsfs}   
\usepackage{calligra}   
\usepackage{url}
\usepackage{hyperref}
\usepackage{threeparttable}
\usepackage{makecell}
\usepackage[normalem]{ulem}
\usepackage{soul}
\usepackage{tikz}
\usepackage[braket, qm]{qcircuit}
\usepackage{booktabs} 
\usepackage{float}

\usepackage{setspace}

\useunder{\uline}{\ul}{}

\begin{document}

\title{PACOX: A FPGA-based Pauli Composer Accelerator for Pauli String Computation}

\author{
	\IEEEauthorblockN{
    Tran Xuan Hieu Le\textsuperscript{1},
    Tuan Hai Vu\textsuperscript{2,3}, 
    Vu Trung Duong Le\textsuperscript{1}, 
    Hoai Luan Pham\textsuperscript{1}, 
    and Yasuhiko Nakashima\textsuperscript{1}}
    
	\IEEEauthorblockA{
    \textsuperscript{1} Nara Institute of Science and Technology, 8916–5 Takayama-cho, Ikoma, Nara 630-0192, Japan.\\
    \textsuperscript{2} University of Information Technology, Ho Chi Minh City, 700000, Vietnam.\\
    \textsuperscript{3} Vietnam National University, Ho Chi Minh City, 700000, Vietnam.\\}
Email: le.tran\_xuan\_hieu.lx0@naist.ac.jp, haivt@uit.edu.vn



}

\markboth{Journal of \LaTeX\ Class Files, Vol. 14, No. 8, August 2015}
{Shell \MakeLowercase{\textit{et al.}}: Bare Demo of IEEEtran.cls for IEEE Journals}
\maketitle

\begin{abstract}
 Pauli strings are a fundamental computational primitive in hybrid quantum-classical algorithms. However, classical computation of Pauli strings suffers from exponential complexity and quickly becomes a performance bottleneck as the number of qubits increases. To address this challenge, this paper proposes the Pauli Composer Accelerator (PACOX), the first dedicated FPGA-based accelerator for Pauli string computation. PACOX employs a compact binary encoding with XOR-based index permutation and phase accumulation. Based on this formulation, we design a parallel and pipelined processing element (PE) cluster architecture that efficiently exploits data-level parallelism on FPGA. Experimental results on a Xilinx ZCU102 FPGA show that PACOX operates at 250 MHz with a dynamic power consumption of 0.33 W, using 8,052 LUTs, 10,934 FFs, and 324 BRAMs. For Pauli strings of up to 19 qubits, PACOX consistently outperforms state-of-the-art CPU-based methods in terms of execution speed, while also requiring significantly less memory and achieving a much lower power–delay product. These results demonstrate that PACOX delivers high computational speed with superior energy efficiency for Pauli-based workloads in hybrid quantum–classical systems.
\end{abstract}


\IEEEpeerreviewmaketitle

\vspace{-0.1cm}
\section{Introduction}

In recent years, the rapid development of hybrid quantum-classical algorithms has intensified the demand for efficient classical processing of quantum operators. Pauli strings are widely used in Hamiltonian simulation, variational quantum algorithms, and quantum error correction. For the development of quantum algorithms, researchers have used software packages such as Qiskit \verb|SparsePauliOp| \cite{qiskit2024} and PennyLane \verb|Pauli| \cite{Bergholm2018PennyLane}. However, in many practical workloads, the dominant computational cost arises from repeated Pauli string computations. At large $n$, these operations quickly become bottleneck due to the exponential growth of the operator.

Numerous studies aim to improve computational speed and support larger values of $n$. The tree-based method in \cite{koska2024tree} reduces arithmetic operations and allows moderately sized problems to run on a single node with limited memory, but it is limited to $n \le 15$. L. Hantzko \textit{et al.} \cite{hantzko2024tensorized} avoid expensive matrix multiplications by computing Pauli weights recursively using matrix slicing, yet their method supports only $n \le 10$. T. Kurita \textit{et al.} \cite{kurita2023pauli} proposed an efficient algorithm to group Pauli strings and reduce measurement cost in variational quantum eigensolver algorithms. Although this approach supports larger $n$, it requires a large amount of memory.

Another emerging trend is the use of graphics processing units (GPUs) for Pauli string computation. In particular, the \verb|apply_pauli_rotation| function in cuQuantum \cite{bayraktar2023cuquantum} is optimized for NVIDIA GPU architectures. In this context, Huang \textit{et al.} \cite{huang2024redefining} proposed a novel approach to improve Pauli string synthesis by incorporating architecture-aware optimization strategies. Although these GPU-based approaches provide significant performance improvements, the authors also reported scalability limitations as $n$ increases, which remains a major challenge for supporting larger quantum systems.

To overcome software limitations, many quantum systems have been developed on hardware platforms, especially using Field-Programmable Gate Arrays (FPGAs). FPGAs are widely used in quantum–classical systems because they support low-latency quantum control, real-time signal processing, and quantum emulation tasks \cite{qin2019fpga,baek2019fpga,vu2025qea}. However, existing FPGA-based approaches primarily focus on quantum control or circuit-level emulation and do not directly address Pauli string computation, which involves exponential index transformations and phase composition. 

To the best of our knowledge, \textbf{this work is the first to propose a dedicated FPGA accelerator for Pauli string computation itself}. This paper proposes PACOX (\textbf{PA}uli \textbf{CO}mposer \textbf{X} Accelerator, where ``X'' is a phonetic shorthand for ``accelerator''), an FPGA-based design for fast Pauli string computation built upon the Pauli Composer (PC) algorithm \cite{paulicomposor}, which addresses the performance limitations of software-based implementations through three key innovations: (1) an optimized memory organization that efficiently manages context data and Pauli matrix storage to support exponential data growth in a hardware-friendly manner, (2) a parallel and pipelined FPGA architecture that exploits data-level parallelism, and (3) a compact binary encoding of Pauli strings that reformulates operator evaluation as XOR-based index permutation and phase accumulation. Together, these design choices enable an efficient and hardware-oriented solution for Pauli string computation.

\section{Improved PACOX Methods} \label{sec:background}

\subsection{Algorithm Formulation}
    \begin{algorithm}[t]
    \small
    \caption{PACOX algorithm}
    \begin{algorithmic}[1]
    \REQUIRE Pauli string $\bm x$
    \STATE $n \leftarrow |\bm x|, L\gets 2^l$
    \STATE $n_Y \leftarrow$ number of $Y$ matrices in $\bm x$
    
    \STATE $\bm k \leftarrow \underbrace{[0,\ldots, 0]}_{2^n},  \bm m \gets \bm k.copy$
    \STATE \textcolor{red}{$\bm v \leftarrow [\tilde{V}[x_{n-1}] \ldots \tilde{V}[x_0]]\ \text{in base 2}$ \COMMENT{$\tilde{V}=[1,1,0,0]$}}
    \STATE $\bm k[0] \leftarrow [\tilde{X}[x_{n-1}] \ldots \tilde{X}[x_0]]\ \text{in base 10}$ \COMMENT{$\tilde{X}=[0,1,1,0]$}
    \STATE $\bm m[0] \leftarrow (-i)^{n_Y \bmod 4} $
    \FOR{$l \in [0, \ldots, n-1]$}
        \STATE \textcolor{red}{[parallel with $N$ $\{\text{PE}_p\}$] \COMMENT{$p = 0, 1, \ldots, N-1$}}
        \STATE \textcolor{red}{$s_p \leftarrow L/N \times p$}
        \STATE \textcolor{red}{$e_p \leftarrow L/N \times (p+1) - 1$}
        \STATE \textcolor{red}{$\bm k[L + s_p : L + e_p] \leftarrow \bm k[s_p : e_p] + (-1)^{\tilde{X}[x[l]]} L$}
    
        \STATE \textcolor{red}{$\bm m[L + s_p : L + e_p]  \leftarrow \bm m[s_p : e_p] \oplus \neg  \bm v[l]$} 
        
    \ENDFOR
    \RETURN A sparse matrix stacking $(\bm k, \bm m)$
    \end{algorithmic}
    \label{algo:ppo}
    \end{algorithm}
    
    The Pauli matrices are Hermitian, Unitary, and together with the identity matrix they form the set $\sigma_{0,1,2,3}=\{I, X, Y, Z\}$. Given an $n$-qubit string $\bm x \in \{0,1,2,3\}^n$, we construct the operator through tensor product ($\otimes$):
    \begin{align}\label{eq:pauli_operator}
        P(\bm x) = \sigma_{x_{n-1}} \otimes \sigma_{x_{n-2}} \otimes \cdots \otimes \sigma_{x_0},
    \end{align}
    
    \noindent which is known as the basic computation in various optimization problems. Let $P_{j,k}(\bm x)$ denote the matrix elements of $P(\bm x)$ ($j,k \in [0,\ldots,2^n-1]$). Native tensor product returns $P(\bm x)$ as $2^n \times 2^n$ matrix, which consumes massive memory for even small $n$. Luckily, the operator $P$ has been proved in \cite{paulicomposor} that for each row $j$, there exists exactly one element at column $\bm k[j]$ such that $P_{j,\bm k[j]}(\bm x) = \pm 1$. Consequently, the problem reduces to determining vector $\bm k$ and the sign of the $2^n$ nonzero entries associated with $\bm x$. The sequential algorithm has been proposed therein; hence, the computational complexity of the original algorithm is lower bounded by $2^n$. Therefore, we present a parallel version in Algorithm~\ref{algo:ppo}, in which the red lines indicate the key improvements introduced over the PC algorithm. At each iteration $l$, the algorithm applies sign inversions and index shifts on $\{\bm k, \bm m\}$, which made the bottleneck of the whole algorithm because their size increases exponentially ($2^n$) based on $n$. Hence, using $N$ data-parallel PEs, the above operations are faster than $N$ times. Note that $N$ must be a power of $2$, then $L/N$ returns the valid index. Furthermore, multiplication operations are replaced by XOR operators, leading to a substantial reduction in computational cycles. In addition, this design choice reduces memory storage requirements by representing data with 2-bit encodings for XOR-based computation, rather than 32-bit values required for multiplication.
    
\subsection{Parallel Execution Strategy}

    \begin{figure}[t]
        \centering
        \includegraphics[width=0.5\textwidth]{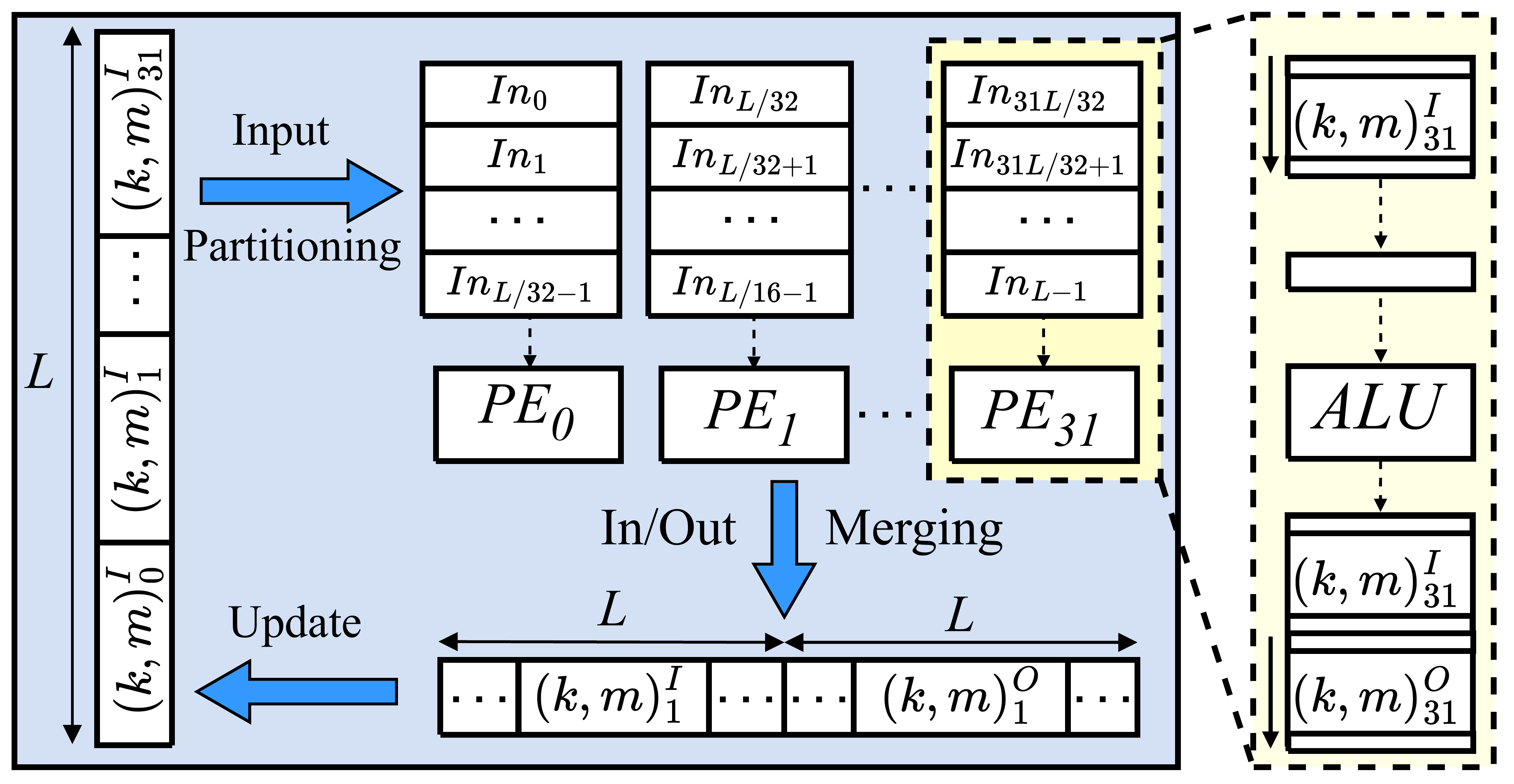}
        \caption{Dataflow and internal datapath of PEs in PACOX method.} 
        \label{fig:data_flow} \vspace{-0.2cm}
    \end{figure}
    \begin{figure}[t]
        \centering
        \includegraphics[width=0.5\textwidth]{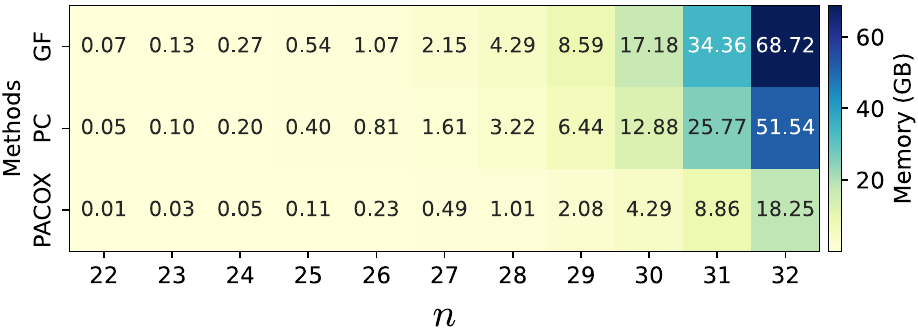}
        \caption{Memory usage between PACOX and PC \cite{paulicomposor}, Gate Fusion (GF) \cite{weko_210570_1} methods.} \vspace{-0.2cm}
        \label{fig:mem_chart} 
    \end{figure}
    
    Fig.~\ref{fig:data_flow} illustrates the data flow of the PACOX algorithm implemented with 32 PEs ($N = 32$). The initial data vectors $(\bm{k}, \bm{m})$, which consist of $L$ tuples $(\bm{k}[i], \bm{m}[i])$, are evenly partitioned into 32 segments. Each segment has a length of $L/32$. After partitioning, the data segments are streamed in a pipelined manner to their corresponding PEs for computation. In each PE, the arithmetic logic unit (ALU) independently updates the assigned $(\bm{k}[i], \bm{m}[i])$ entries. The computed results are stored in local memory, which also holds the original input data. The updated results are then merged with the existing data to form the input for the next computation cycle, referred to as \emph{In/Out Merging}. By exploiting data-level parallelism with 32 PEs, repeated for $l$ iterations, the execution time of this computation is reduced by approximately a factor of 32 compared to the sequential implementation.
   
\vspace{-0.2cm}
\subsection{Memory Complexity Analysis}
    
    Fig.~\ref{fig:mem_chart} illustrates the theoretical relationship between the $n$ and memory usage (GB) across different memory storage methods. As shown, the memory requirements for Pauli string computation grow exponentially with increasing $n$, which poses a major challenge for large-scale problems. For smaller problem sizes, memory consumption remains modest. For example, when $n = 22$, all evaluated methods require less than 1~GB of memory. However, as $n$ increases, clear differences emerge among the methods. At $n = 32$, the memory usage of the PC and GF methods rises sharply to approximately 51.54 GB and 68.72 GB, respectively. In contrast, PACOX requires only 18.25 GB at the same problem size. This significant reduction is achieved by storing only the nonzero elements of Pauli matrices and by employing a compact binary encoding. These results demonstrate that PACOX not only improves computational performance but also substantially reduces memory requirements, making it more scalable for large Pauli string computations.

    \begin{figure*}[t]
        \centering
        \includegraphics[width=1\textwidth]{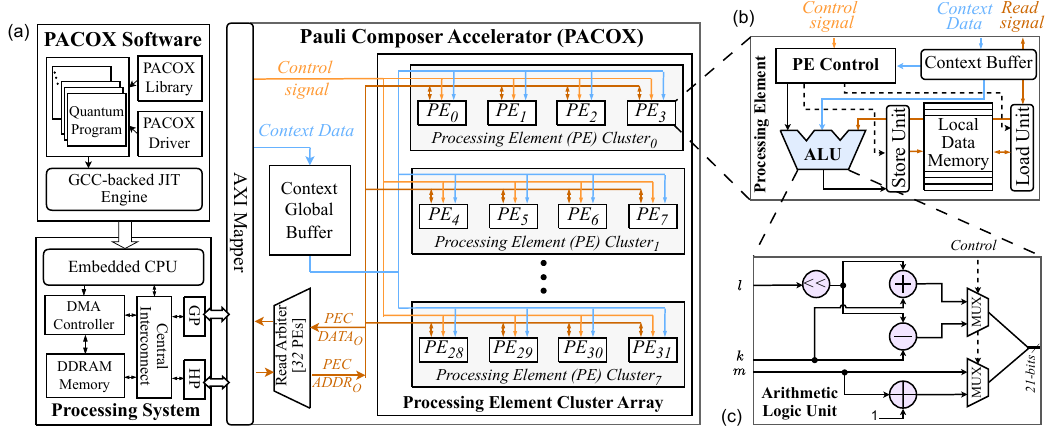}
        \caption{The detailed hardware architecture of PACOX: (a) Overview of PACOX architecture, (b) Processing Element, and (C) Arithmetic Logic Unit.} 
        \label{fig:overview}
    \end{figure*}
\section{Hardware Architecture} \label{sec:related_work}

\subsection{Overview}

Fig.~\ref{fig:overview}a illustrates the system-level architecture of the proposed PACOX implemented on a system-on-chip (SoC) platform. The architecture consists of two main components: the processing system (PS) and the programmable logic (PL). The PS, which is divided into user space and kernel space, runs the GNU/Linux operating system and is responsible for application-level control. In the user space, programmers construct Context data structures that describe the corresponding Pauli strings. These contexts are compiled in the kernel space by a GCC-based PACOX compiler and transferred to the PL via the AXI bus using programmable input/output (PIO). Since control signals are lightweight and configuration data are transmitted only once, both are efficiently delivered using a 64-bit PIO interface. The PACOX hardware in the PL consists of four main modules: an AXI Mapper, a Context Global Buffer, a Read Arbiter, and a PE Cluster (PEC) Array. The AXI Mapper manages data communication between the PS and the local data memory of the PEs. 
When the Context Global Buffer receives the configuration data from the PS, the PEC Array starts computing the vectors $(\bm k,\bm m)$. Once the computation is finished, the PEC Array signals the PS, and the results $\bm{k}$ and $\bm{m}$ are fetched into DDR memory via the AXI Mapper and Read Arbiter using DMA.

\subsection{High-performance Processing Element and ALU}
    
    To match the 128-bit width of the DMA bus, four PEs are grouped into a single cluster, enabling the simultaneous readout of four output data items per transfer. Based on the available resources of the ZCU102 platform, the PACOX architecture is configured with a total of 32 PEs organized into a PEC array, corresponding to eight PECs. Fig.~\ref{fig:overview}b illustrates the architecture of an individual PE. Each PE incorporates a Context Buffer to reduce fan-in and fan-out overhead during data transmission from PS to PE control logic. In addition to the basic logic blocks responsible for synchronization and pipelined computation across PEs, a Local Data Memory (LDM) is introduced. Each LDM stores a vector of size $2^{14}$, where each entry is a tuple $(\bm{k}[i],\bm{m}[i])$. Here, $\bm{k}[i]$ is a 19-bit integer, and $\bm{m}[i]$ is a 2-bit integer encoding values from $0$ to $3$, corresponding to $[1,\,-1,\,1j,\,-1j]$. Fig. \ref{fig:overview}c shows the architecture of the ALU. It accepts three inputs, namely $\bm{k}$, $\bm{m}$, and $\bm{l}$, and produces an output tuple $(\bm{k}, \bm{m})$. In this design, conventional multiplication operations are replaced by XOR-based logic enabled by the proposed binary encoding of Pauli strings. This replacement significantly reduces computational complexity and hardware cost compared with arithmetic multiplication. Owing to the simplicity of the ALU design, all computations are completed within a single clock cycle.

\subsection{Configurable Memory Organization}
    \begin{figure}[t]
        \centering
        \includegraphics[width=0.45\textwidth]{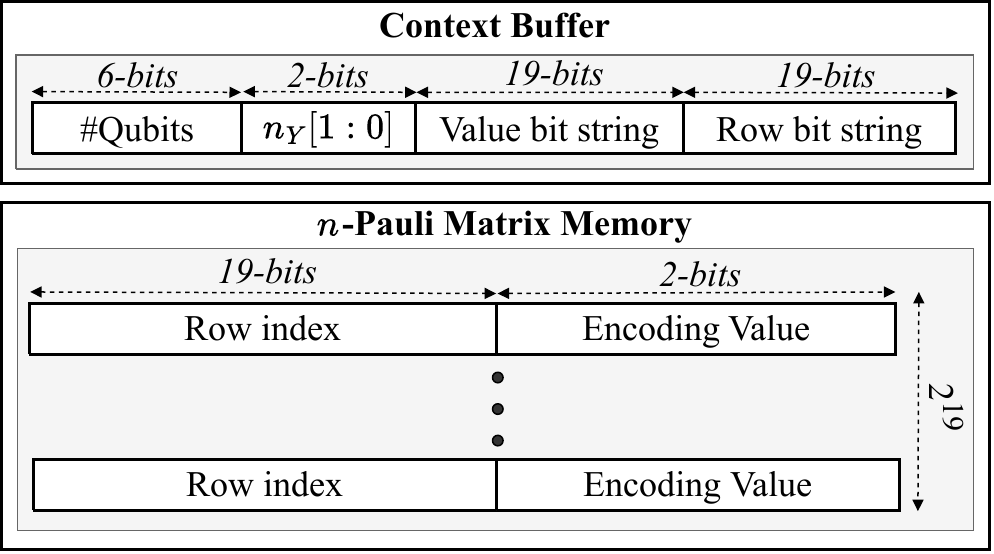}
        \caption{Memory organization of PACOX.} 
        \label{fig:memory} 
    \end{figure}
    
    As discussed in Section~\ref{sec:background}, the matrix size grows exponentially with the $n$, while FPGA platforms are limited by on-chip memory resources. Fig.~\ref{fig:memory} shows the memory organization of the proposed PACOX design for efficient FPGA implementation. Two types of memory are used: Context Buffers and the $n$-Pauli Matrix Memory. The Context Buffers store context data and include a 6-bit signal that indicates the number of qubits supported by PACOX. In addition, the Context Buffer stores other configuration parameters, such as $n_Y[1{:}0]$, the Row bit string, and the Value bit string. The Row bit string is defined as $[\tilde{X}[x_{n-1}] \ldots \tilde{X}[x_0]]_{2}$, while the Value bit string is given by $[\tilde{V}[x_{n-1}] \ldots \tilde{V}[x_0]]_{2}$. The $n$-Pauli Matrix Memory is the largest memory component and is used to store the vectors $(\bm k, \bm m)$. To support up to $n = 19$, this memory requires a depth of $2^{19}$. When the $n$-Pauli Matrix Memory is implemented on FPGA platforms with more abundant BRAM resources, such as the Xilinx Alveo or Versal series, PACOX can be scaled to support larger values of $n$.

\section{Experiments and Results} \label{sec:results}
\subsection{Analysis of Post-Implementation Result} \label{subsec:resource}
    To realize the proposed PACOX architecture shown in Fig.~\ref{fig:overview}, the design was synthesized and implemented on the Xilinx ZCU102 FPGA platform using Vivado Design Suite version 2021.2. The synthesis and implementation processes employed the \texttt{Flow\_areaOptimized\_high} and \texttt{Performance\_ExtraTimingOpt} strategies, respectively, to achieve a balanced trade-off between logic utilization, timing closure, and power efficiency. Tab.~\ref{tab:resources} summarizes the post-implementation resource utilization and power consumption of the PACOX design, with a breakdown across its major functional modules, including the AXI Mapper, Read Arbiter, Context Buffer, and Cluster Array.

    \begin{table}[H]
    \centering
    \caption{PACOX DESIGN UTILIZATION ON THE ZCU102 FPGA (POST-IMPLEMENTATION).}
    \footnotesize
    \label{tab:resources}
    \renewcommand{\arraystretch}{1.2}
    \resizebox{1\linewidth}{!}{%
    \begin{tabular}{ccccc c}
    \toprule
    \multirow{2}{*}{\textbf{Design Name}} &
    \multirow{2}{*}{\textbf{\begin{tabular}[c]{@{}c@{}}Freq.\\ (MHz)\end{tabular}}} &
    \multicolumn{3}{c}{\textbf{Resources}} &
    \multirow{2}{*}{\textbf{\begin{tabular}[c]{@{}c@{}}Power\\ (W)\end{tabular}}} \\
    \cmidrule(lr){3-5}
     &  & \textbf{LUTs} & \textbf{FFs} & \textbf{BRAM} &  \\
    \midrule
    AXI Mapper      & \multirow{5}{*}{250} & 322  & 252   & 0   & 0.005  \\
    Read Arbiter    &                      & 127  & 574   & 0   & 0.001  \\
    Context Buffer  &                      & 0    & 1,886  & 0   & 0.0292 \\
    Cluster Array   &                      & 7,603 & 8,222  & 324 & 0.2948 \\
    \midrule
    \textbf{PACOX (Total)} & \textbf{250}
     & \textbf{8,052} & \textbf{10,934} & \textbf{324} & \textbf{0.33} \\
    \bottomrule
    \end{tabular}}
    \end{table}
    
  The Cluster Array dominates overall resource usage due to its computational intensity and parallel processing structure, accounting for the majority of LUTs, FFs, and Block RAMs. In contrast, control-oriented modules such as the AXI Mapper and Read Arbiter incur minimal hardware overhead, demonstrating the effectiveness of the modular and hierarchical design. Overall, the complete PACOX architecture occupies 8,052 LUTs, 10,934 FFs, and 324 Block RAM while operating at 250~MHz. Despite the high operating frequency and the use of on-chip memory resources, the total power consumption remains limited to 0.33~W, indicating that the design can be integrated with additional processing modules without exceeding the device constraints of the target FPGA platform.
  
    \begin{figure}[t]
        \centering
        \includegraphics[width=0.4\textwidth]{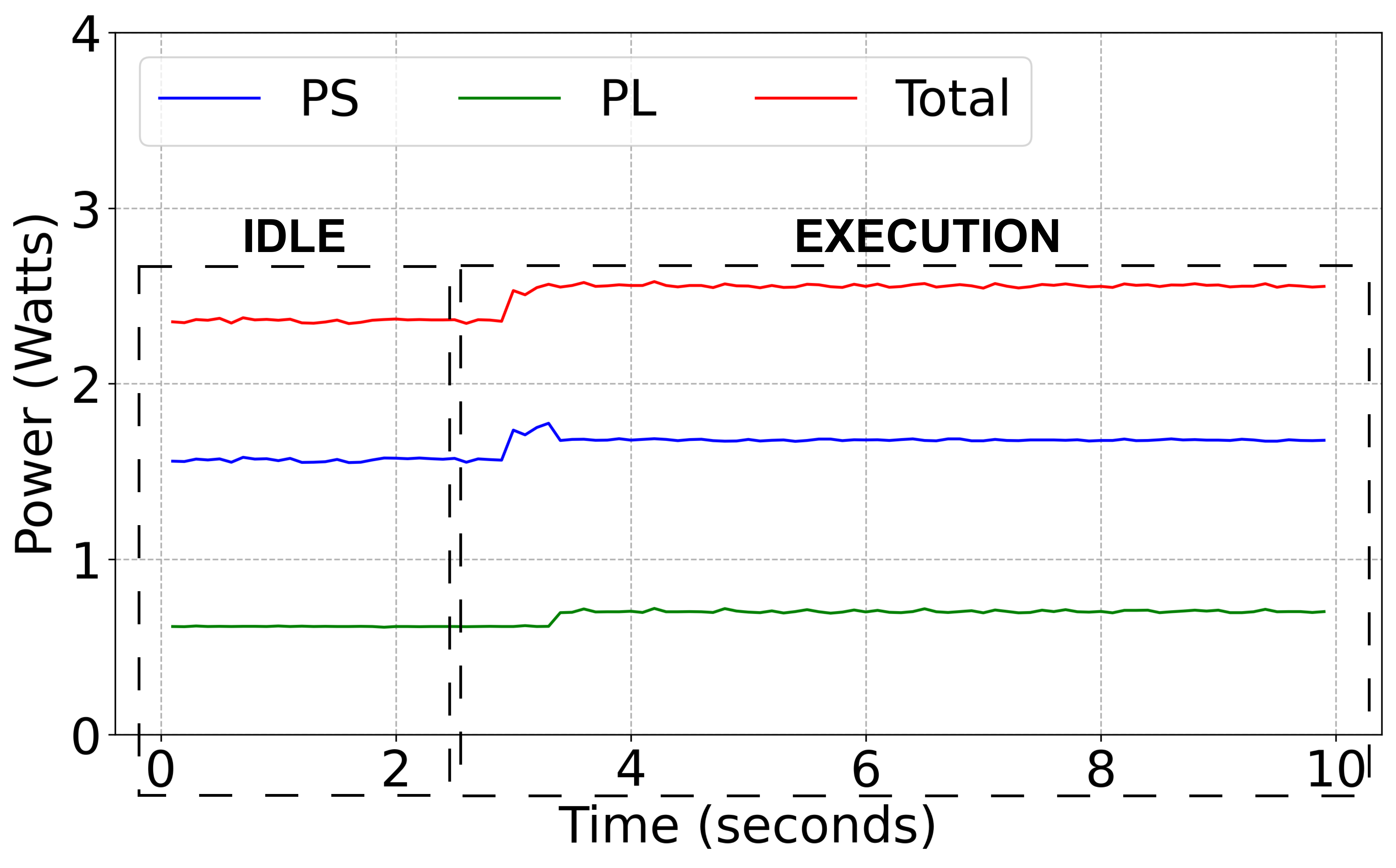}
        \caption{Power measurement of PACOX on random $n$-Pauli strings.}
        \label{fig:power}  
    \end{figure}
    \begin{figure}[t]
        \centering
        \includegraphics[width=0.5\textwidth]{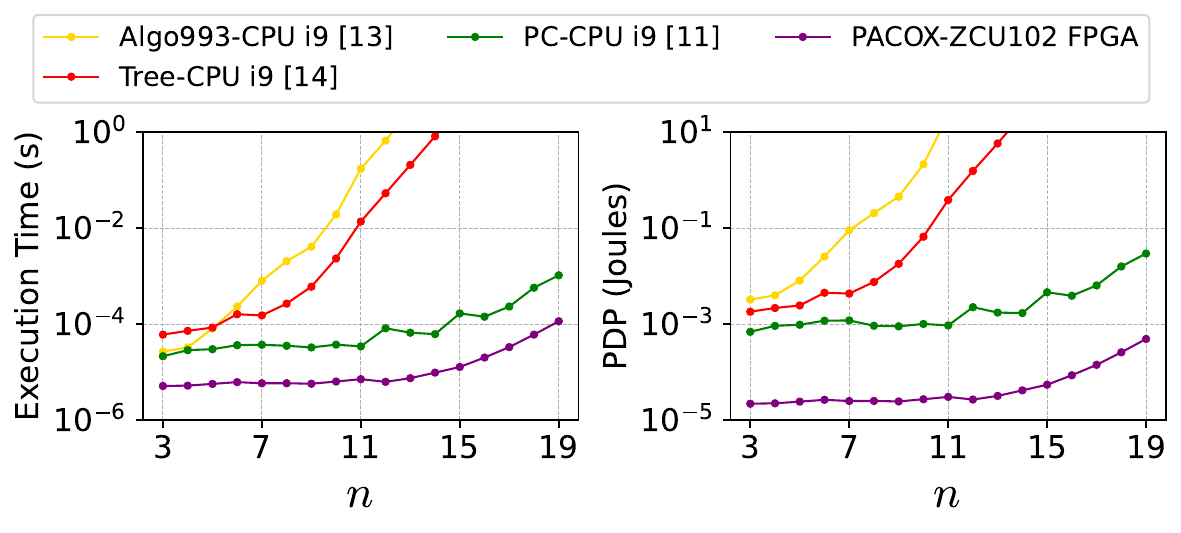}
        \caption{Comparison of Execution Time and PDP for random $n$-qubit Pauli strings: PACOX vs. PC \cite{paulicomposor}, Algo993 \cite{fackler2019algorithm} and Tree \cite{482008} methods.}
        \label{fig:time_cpu} \vspace{-0.2cm}
    \end{figure}
    
    \begin{figure*}[t]
        \centering
        \includegraphics[width=1\textwidth]{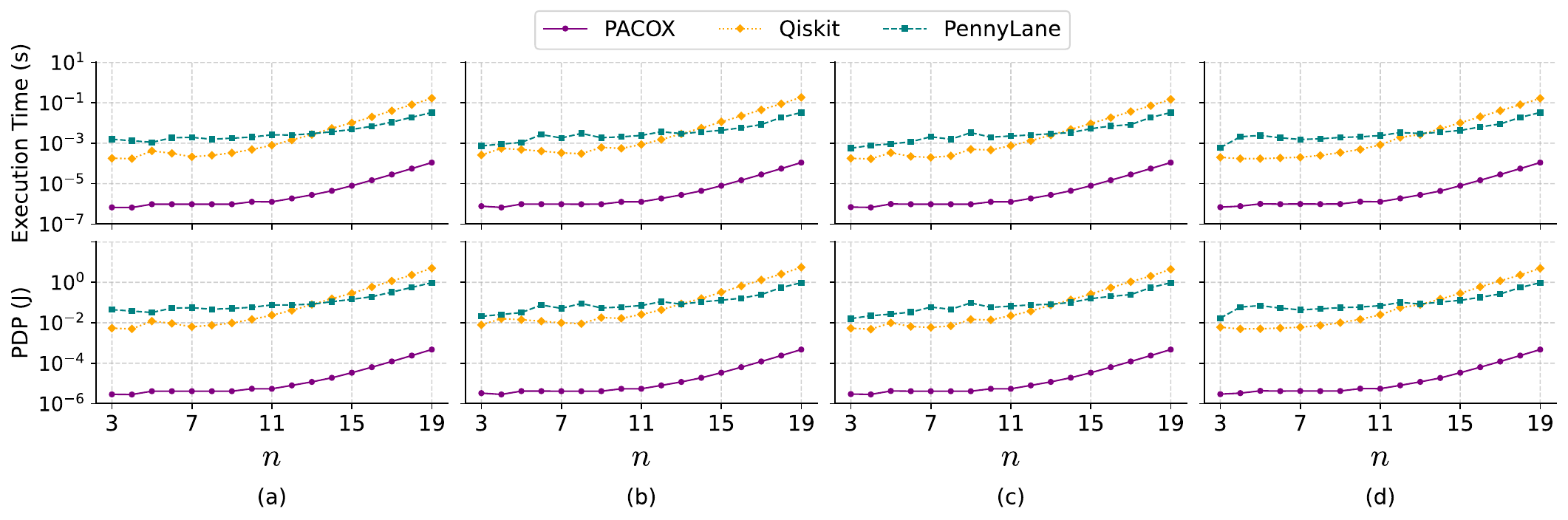}
        \caption{Execution times and PDP comparison between PACOX and CPU-based quantum software libraries (Qiskit and PennyLane) across the (a) Heisenberg \cite{nielsen2010quantum}, (b) QChem \cite{peruzzo2014variational}, (c) TFIM \cite{Sachdev_2011} and (d) Stabilizer \cite{gottesman1997stabilizer} testcases.} \vspace{-0.2cm}
        \label{fig:time_software}  
    \end{figure*}
    \begin{figure}[t]
        \centering
        \includegraphics[width=0.5\textwidth]{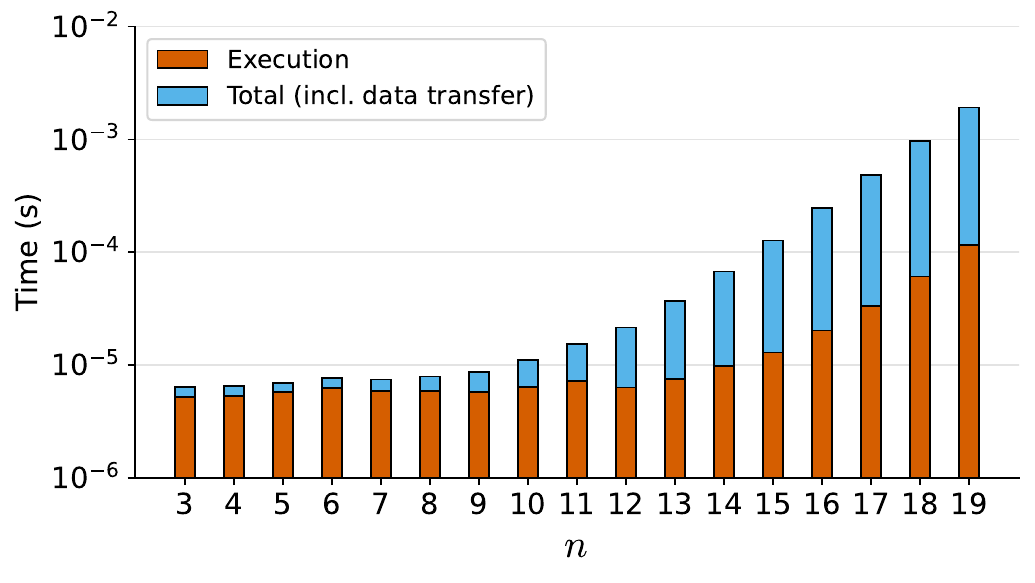}
        \caption{Runtime Analysis of PACOX with Varying Qubit Counts.}
        \label{fig:time_total} \vspace{-0.2cm}
    \end{figure}
    
\subsection{Real-Time Power Measurement}
    Following the post-implementation analysis presented in Section \ref{subsec:resource}, we further evaluate its runtime energy behavior through on-board power measurements. To accurately capture real-time power consumption during execution, the INA226 power monitoring sensor integrated on the ZCU102 FPGA platform was employed. This measurement provides a dynamic validation of the synthesis and implementation results by reflecting the actual power usage under representative workloads.

    The measured power profiles are illustrated in Fig. \ref{fig:power} for the execution of random $n$-Pauli strings, where $n = 3,4, \dots ,19$. The figure reports the power consumption of the PS, PL, and the total system power over time, clearly distinguishing between idle and execution phases. During the idle state, the power consumption remains relatively stable, whereas a noticeable increase is observed once the PACOX computation is triggered, confirming the activation of the PL accelerator and associated data movement.
    Specifically, the PS power ranges from 1.55 W to 1.77 W, corresponding to approximately 0.22 W of dynamic power, while the PL consumes between 0.61 W and 0.72 W during execution. The total system power reaches a maximum of 2.58 W, with an associated dynamic power of 0.24 W. Consequently, the effective power attributed to PACOX execution can be calculated by combining the maximum PL power with the total dynamic power, reaching 0.96 W. This result highlights the energy efficiency of the proposed design and confirms that the PACOX accelerator can sustain high-throughput operation with limited power overhead.

\subsection{Comparison with State-of-the-Art Methods}

    Fig.~\ref{fig:time_cpu} compares the execution time and power--delay product (PDP) of different computing platforms for $n = 3,4,\dots,19$. All CPU experiments are conducted on an Intel i9-10940X processor operating at 3.30~GHz. Across the evaluated range, PACOX consistently outperforms all CPU-based approaches in terms of execution time. Compared with the PC algorithm, PACOX achieves speedups ranging from \textbf{4.15 times} ($5.18\times10^{-6}$ vs. $21.6\times10^{-6}$~s) to \textbf{9 times} ($1.15\times10^{-4}$ vs. $10.44\times10^{-4}$~s). The performance advantage becomes even more pronounced when compared with the Algo993~\cite{fackler2019algorithm} and Tree~\cite{482008} methods, where the speedup increases from \textbf{11.7 times} ($5.18\times10^{-6}$ vs. $60.8\times10^{-6}$~s) up to $\mathbf{2 \times} \mathbf{10^{6}}$ \textbf{times} ($1.15\times10^{-4}$ s vs. 231.92~s). These results clearly indicate that PACOX substantially reduces execution time compared with state-of-the-art CPU-based methods, and that its performance advantage grows rapidly as the number of qubits increases.

    In addition to reducing execution time, PACOX also significantly outperforms CPU-based methods in terms of PDP. When compared with the PC method, PACOX achieves PDP improvements ranging from \textbf{3.14 times} ($2.22\times10^{-5}$ vs. $6.96\times10^{-5}$~J) to \textbf{60.1 times} ($4.93\times10^{-4}$ vs. $2.96\times10^{-2}$~J) as the number of qubits increases. The advantage of PACOX becomes even more pronounced when compared with the Algo993 and Tree methods. In these cases, the PDP improvement increases from \textbf{147.11 times} ($2.22\times10^{-5}$ vs. $3.25\times10^{-3}$~J) up to $\mathbf{6.06 \times 10^{7}}$ \textbf{ times} ($4.93\times10^{-4}$ vs. $2.99\times10^{4}$~J). This dramatic gap is mainly due to the exponential growth in computation time and energy consumption of conventional CPU-based approaches, while PACOX effectively mitigates this effect through compact data representation and highly parallel hardware execution. Consequently, PACOX represents a promising solution for large-scale Pauli string computation, particularly in hybrid quantum--classical systems where power and performance are critical constraints.
    
\vspace{-0.2cm}
\subsection{Comparison with CPU-Based Software}\label{subsec:time_cpu}
 
    We further compare the proposed PACOX with widely used CPU-based quantum software libraries on standard datasets, including Qiskit \cite{qiskit2024} and PennyLane \cite{Bergholm2018PennyLane} conducted on an Intel 3.30~GHz i9-10940X processor. Fig. ~\ref{fig:time_software} shows the execution time for computing $n$-Pauli strings on the Heisenberg \cite{nielsen2010quantum}, QChem \cite{peruzzo2014variational}, TFIM \cite{Sachdev_2011}, and Stabilizer \cite{gottesman1997stabilizer} testcases with qubit counts ranging from 3 to 19. All results correspond to real executions on both platforms. Across all datasets, PACOX consistently outperforms Qiskit and PennyLane, and the performance gap widens as the number of qubits increases. This highlights the effectiveness of the proposed hardware architecture for accelerating practical Pauli-based workloads. Fig.~\ref{fig:time_software} also reports the measured PDP for PACOX and the CPU-based libraries over the same qubit range. PACOX achieves significantly lower PDP values across all datasets, demonstrating that its performance gains are obtained with superior energy efficiency.

    Fig. \ref{fig:time_total} shows the total runtime of the PACOX system for varying values of $n$ from 3 to 19. As the $n$ increases, both execution time and total runtime (including data transfer time) increase. However, the data transfer time grows much more significantly than the execution time. This happens because PACOX uses 32 PEs to compute 32 input data points simultaneously. But, the ZCU102 FPGA's DMA can only transfer 4 output data points at a time from the PL to the PS. Specifically, the bandwidth of the ZCU102 FPGA is only 4 GB/s working at a frequency of 250 MHz. Additional time is also added due to tasks such as processing communications and sending initial signals from the PS to the PL. As a result, the total runtime increases much more than the execution time. This bottleneck in data transfer could be alleviated by using the Alveo U280 FPGA, which offers high-bandwidth memory of up to 64 GB/s at a frequency of 250 MHz.

\vspace{-0.2cm}
\section{Conclusion}\label{sec:conclusion}

This paper presents PACOX, a dedicated FPGA-based architecture for accelerating Pauli string computation in hybrid quantum-classical systems. PACOX employs a compact binary encoding that transforms the entire process into an XOR-based index permutation and phase accumulation, enabling efficient parallel and pipelined execution on an FPGA. Implemented on a Xilinx ZCU102 platform, PACOX outperform representative CPU-based Pauli computation methods, while significantly reducing memory usage and power consumption. These results demonstrate that PACOX provides a high-speed and energy-efficient hardware solution for accelerating Pauli-based quantum workloads.

\section*{Acknowledgment}
This work was supported by JST-ALCA-Next (JPMJAN23F4), the NAIST Senju Monju Project (Daiichi-Sankyo ”Habataku” Support Program for the Next Generation of Researchers).

\section*{Appendix: Use case of PACOX}
In this section, we introduce several problems which are benchmarked in Fig.~\ref{fig:time_software}.

\noindent\textbf{Two-local Pauli strings.} These operators describe nearest-neighbor interactions such as $XX$, $YY$, and $ZZ$ in many fundamental Hamiltonians, including spin-chain models used in variational quantum algorithms. They are expressed as
\begin{align}\label{eq:two-local}
    P^{x}_{j}
    = I^{\otimes j} \otimes \sigma_x \sigma_x \otimes I^{\otimes (n-j-2)},
    \quad x \in \{X, Y, Z\}. \nonumber
\end{align}

\noindent\textbf{Stabilizer Code.} The Pauli string
$Z_j = I^{\otimes j} \otimes Z \otimes I^{\otimes (n-j-1)}$
is a basic element of the stabilizer generator
$\mathbb{P} = \{Z_0, \ldots, Z_n\}$.
The observable $\langle Z_0 \rangle$ also serves as a basic measurement observable.

\noindent\textbf{Variational Quantum Eigensolver (VQE).}
We use the Hamiltonian of the LiH molecule with atomic coordinates
$[[0,0,0],[0,0,1.6\,\text{\AA}]]$ in the STO-3G basis. The Pauli strings are expressed as the sum 
\begin{align}
    H_{\text{LiH}}
    = -7.3147 I^{\otimes 4} + \ldots + 0.0843 Z_2 Z_3. \nonumber
\end{align}

\noindent\textbf{Transverse-Field Ising Model (TFIM).}
The TFIM is a paradigmatic quantum spin model that captures the competition between nearest-neighbor Ising interactions and a transverse magnetic field. Its Hamiltonian is also given by the sum of Pauli strings
\begin{align}
    H_{\text{Ising}}
    = -J \sum_{j=1}^{N-1} Z_j Z_{j+1}
      - h \sum_{j=1}^{N} X_j, \nonumber
\end{align}

\noindent
where $Z_j Z_{j+1}$ represents the nearest-neighbor Ising interaction, and $X_j$ denotes the transverse-field term. In this work, the external field $h$ is turned on.

We measure the evaluation time from initialization until obtaining the sparse matrix stacking of
$\{P^{x}_{j}\}$, $\mathbb{P}$, $H_{\text{LiH}}$, or $H_{\text{Ising}}$.
For larger problem instances, we increase the number of qubits or spin orbitals.

\bibliographystyle{IEEEtran}
\bibliography{references.bib}

@article{paulicomposor,
author={Vidal Romero, Sebasti{\'a}n
and Santos-Su{\'a}rez, Juan},
title={{PauliComposer: compute tensor products of Pauli matrices efficiently}},
journal={Quantum Information Processing},
year={2023},
month={Dec},
day={15},
volume={22},
number={12},
pages={449},
issn={1573-1332},
doi={10.1007/s11128-023-04204-w},
}

@ARTICLE{482008,
  author={Selesnick, I.W. and Burrus, C.S.},
  journal={IEEE Transactions on Signal Processing}, 
  title={{Automatic generation of prime length FFT programs}}, 
  year={1996},
  volume={44},
  number={1},
  pages={14-24},
  doi={10.1109/78.482008}}

@article{fackler2019algorithm,
  title={{Algorithm 993: Efficient computation with kronecker products}},
  author={Fackler, Paul L},
  journal={ACM Transactions on Mathematical Software (TOMS)},
  volume={45},
  number={2},
  pages={1--9},
  year={2019},
  publisher={ACM New York, NY, USA}
}

@techreport{weko_210570_1,
   author	 = "Hiroshi,Horii and Jun,Doi",
   title	 = "Optimization of Quantum Computing Simulation with Gate Fusion",
   year 	 = "2021",
   institution	 = "IBM Quantum, IBM Research Tokyo, IBM Quantum, IBM Research Tokyo",
   number	 = "23",
   month	 = "mar"
}

@book{nielsen2010quantum,
  title={{Quantum computation and quantum information}},
  author={Nielsen, Michael A and Chuang, Isaac L},
  year={2010},
  publisher={Cambridge university press}
}

@article{peruzzo2014variational,
  title={{A variational eigenvalue solver on a photonic quantum processor}},
  author={Peruzzo, Alberto and McClean, Jarrod and Shadbolt, Peter and Yung, Man-Hong and Zhou, Xiao-Qi and Love, Peter J and Aspuru-Guzik, Al{\'a}n and O’brien, Jeremy L},
  journal={Nature communications},
  volume={5},
  number={1},
  pages={4213},
  year={2014},
  publisher={Nature Publishing Group UK London}
}

@book{Sachdev_2011, 
place={Cambridge}, 
edition={2},
title={{Quantum Phase Transitions}}, 
publisher={Cambridge University Press}, 
author={Sachdev, Subir}, 
year={2011}
}

@book{gottesman1997stabilizer,
  title={{Stabilizer codes and quantum error correction}},
  author={Gottesman, Daniel},
  year={1997},
  publisher={California Institute of Technology}
}

@misc{qiskit2024,
      title={Quantum computing with {Q}iskit},
      author={Javadi-Abhari, Ali and Treinish, Matthew and Krsulich, Kevin and Wood, Christopher J. and Lishman, Jake and Gacon, Julien and Martiel, Simon and Nation, Paul D. and Bishop, Lev S. and Cross, Andrew W. and Johnson, Blake R. and Gambetta, Jay M.},
      year={2024},
      doi={10.48550/arXiv.2405.08810},
      eprint={2405.08810},
      archivePrefix={arXiv},
      primaryClass={quant-ph}
}

@article{bergholm2018pennylane,
  title={Pennylane: Automatic differentiation of hybrid quantum-classical computations},
  author={Bergholm, Ville and Izaac, Josh and Schuld, Maria and Gogolin, Christian and Ahmed, Shahnawaz and Ajith, Vishnu and Alam, M Sohaib and Alonso-Linaje and others},
  journal      = {arXiv preprint},
  volume       = {arXiv:1811.04968},
  year         = {2018},
}

@inproceedings{bayraktar2023cuquantum,
  title={{cuQuantum SDK: A High-Performance Library for Accelerating Quantum Science}},
  author={Bayraktar, Harun and Charara, Ali and Clark, David and Cohen, Saul and Costa, Timothy and Fang, Yao-Lung L and Gao, Yang and Guan, Jack and Gunnels, John and Haidar, Azzam and others},
  booktitle={2023 IEEE International Conference on Quantum Computing and Engineering (QCE)},
  volume={1},
  pages={1050--1061},
  year={2023},
  organization={IEEE}
}

@inproceedings{koska2024tree,
  title={{A tree-approach Pauli decomposition algorithm with application to quantum computing}},
  author={Koska, Oc{\'e}ane and Baboulin, Marc and Gazda, Arnaud},
  booktitle={ISC High Performance 2024 Research Paper Proceedings (39th International Conference)},
  pages={1--11},
  year={2024},
  organization={Prometeus GmbH}
}

@article{hantzko2024tensorized,
  title={{Tensorized Pauli decomposition algorithm}},
  author={Hantzko, Lukas and Binkowski, Lennart and Gupta, Sabhyata},
  journal={Physica Scripta},
  volume={99},
  number={8},
  pages={085128},
  year={2024},
  publisher={IOP Publishing}
}

@article{kurita2023pauli,
  title={{Pauli string partitioning algorithm with the Ising model for simultaneous measurements}},
  author={Kurita, Tomochika and Morita, Mikio and Oshima, Hirotaka and Sato, Shintaro},
  journal={The Journal of Physical Chemistry A},
  volume={127},
  number={4},
  pages={1068--1080},
  year={2023},
  publisher={ACS Publications}
}

@inproceedings{huang2024redefining,
  title={{Redefining Lexicographical Ordering: Optimizing Pauli String Decompositions for Quantum Compiling}},
  author={Huang, Qunsheng and Winderl, David and Meijer-Van De Griend, Arianne and Yeung, Richie},
  booktitle={2024 IEEE International Conference on Quantum Computing and Engineering (QCE)},
  volume={1},
  pages={885--896},
  year={2024},
  organization={IEEE}
}

@article{baek2019fpga,
  title={{An FPGA-based quantum feedback system for real-time qubit control}},
  author={Baek, Unpil and Xu, Yilun and Huang, Gang and Doolittle, Lawrence and Siddiqi, Irfan},
  journal={Bulletin of the American Physical Society},
  volume={64},
  year={2019},
  publisher={APS}
}

@article{qin2019fpga,
  title={{An FPGA-based hardware platform for the control of spin-based quantum systems}},
  author={Qin, Xi and Zhang, Wenzhe and Wang, Lin and Zhao, Yuxi and Tong, Yu and Rong, Xing and Du, Jiangfeng},
  journal={IEEE Transactions on Instrumentation and Measurement},
  volume={69},
  number={4},
  pages={1127--1139},
  year={2019},
  publisher={IEEE}
}

@inproceedings{vu2025qea,
  title={{QEA: An Accelerator for Quantum Circuit Simulation with Resources Efficiency and Flexibility}},
  author={Vu, Tuan Hai and Le, Vu Trung Duong and Pham, Hoai Luan and Nakashima, Yasuhiko and others},
  booktitle={2025 10th IEEE International Conference on Integrated Circuits, Design, and Verification (ICDV)},
  pages={55--60},
  year={2025},
  organization={IEEE}
}

\end{document}